# Polarization effects on the dielectric properties of molten AgI

**V Bitrián and J Trullàs**

Departament de Física i Enginyeria Nuclear, Universitat Politècnica de Catalunya, Campus Nord UPC, 08034 Barcelona, Spain

E-mail: vicente.bitrian@upc.edu

**Abstract**. The results are reported of molecular dynamics simulations of the static longitudinal dielectric and response functions for molten AgI at 923 K using two ionic models. The first one is a rigid ion model, while in the second the induced dipole moments of the anions are added to the effective pair potentials of the first. It is derived theoretically that the dielectric functions for the polarizable ion model are determined by spatial correlations of charge and dipole moment densities. The charge structure factor at long wavelengths is also studied.

## 1. Introduction

Recently, it has been shown from molecular dynamics (MD) simulations that the main trends of the experimental static structure factor of molten AgI at 923 K may be reproduced by using a polarizable ion model (PIM) in which induced point dipoles are added on the anionic sites [1]. The inclusion of the anions polarizability to simple effective rigid ion pair potentials of the form proposed by Parrinello *et al*. [2] for *α*-AgI accounts for the experimental prepeak at about 1 Å$^{-1}$ (see figure 1). The structural origin of the prepeak has been related to a length scale, larger than the distance between neighbouring ions, which characterizes the ordering of the voids between cations. Further analysis of these MD results shows that the PIM values of the charge structure factor $S_{ZZ}(k)$ at low wave numbers are higher than the theoretical long-wavelength limit approximation derived for rigid ion models [3] (see figure 2). This result, since $S_{ZZ}(k)$ is directly related to the static longitudinal dielectric function of rigid ion models, prompted us to study this function for the polarizable case.

The static longitudinal dielectric function $\epsilon_L(k)$ determines the linear response of the system to a weak external field, and it has been calculated for a wide variety of systems using models that include only point charges [4] or only point dipoles [5], but not both as in [1]. In this paper we show that, for polarizable ion models in which both point charges and dipoles are present, $\epsilon_L(k)$ results from response function contributions related to charge-charge, dipole-dipole and charge-dipole correlations.

## 2. Ionic models

In the present work we have simulated a rigid ion model (RIM) of AgI with effective pair potentials of the form proposed by Parrinello *et al* [2],

$$f_{ab}(r) = f^0_{ab}(r) - \frac{P_{ab}}{r^4}, \qquad \text{with} \qquad f^0_{ab}(r) = \frac{z_a z_b e^2}{r} + \frac{H_{ab}}{r^{h_{ab}}} - \frac{C_{ab}}{r^6}. \qquad (1)$$

The first term of $f^0_{ab}(r)$ is the Coulomb interaction between charges, with $z_a < 1$ the effective charge in units of the fundamental charge *e*; the second models the repulsion between the ions, and the third is the van der Waals contribution. The second term of $f_{ab}(r)$ denotes the effective monopole-induced dipole interaction with $P_{ab} = (1/2)(a_a z_b^2 + a_b z_a^2)e^2$, where $a_a$ are the polarizabilities. This is the way in which polarization is approximated in the RIM, ignoring the many-body nature of induced polarization. For the parameterization of $f_{ab}(r)$ we have chosen the values given by Shimojo and Kobayashi [6], who reproduced the α-phase properties of AgI at the appropriate thermodynamic state. These values are $|z| = 0.5815$, $a_- = 6.12$ Å$^3$ and $a_+ = 0$. Then, $P_{++} = 0$ and $P_{--} = 2P_{+-} = 29.796$ eVÅ$^4$. The other parameters are $h_{++} = 11$, $h_{+-} = 9$, and $h_{--} = 7$; $H_{++} = 0.162$ eVÅ$^{11}$, $H_{+-} = 1309.1$ eVÅ$^9$, and $H_{--} = 5325.2$ eVÅ$^7$; $C_{--} = 84.4$ eVÅ$^6$ and $C_{++} = C_{+-} = 0$.

The polarizable ion model (PIM) of AgI is constructed by adding he induced polarization contributions to $f_{ab}^0(r)$. We assume that, on an ion placed at position $\mathbf{r}_i$, the local electric field $\mathbf{E}_i$ due to all the other ions induces a dipole whose moment is $\mathbf{p}_i = a_i \mathbf{E}_i$. Then, the local field is $\mathbf{E}_i = \mathbf{E}_i^q + \mathbf{E}_i^p$, where $\mathbf{E}_i^q$ is the field at $\mathbf{r}_i$ due to all the point charges except $q_i = z_i e$ at $\mathbf{r}_i$, and $\mathbf{E}_i^p$ the field at $\mathbf{r}_i$ due to all the dipole moments except $\mathbf{p}_i$ [7]. The potential energy of this polarizable model may be written as

$$U = \frac{1}{2}\sum_{i=1}^{N}\sum_{j\neq i}^{N} f_{ij}^0(r_{ij}) - \sum_{i=1}^{N}\mathbf{p}_i \cdot \mathbf{E}_i^q - \frac{1}{2}\sum_{i=1}^{N}\mathbf{p}_i \cdot \mathbf{E}_i^p + \sum_{i=1}^{N}\frac{\mathbf{p}_i^2}{2a_i}. \tag{2}$$

## 3. Dielectric response theory

As it is well known, the static linear charge response function of systems of rigid ions to an external electric field is determined by the charge density autocorrelation function [3]. The static linear polarization response function of systems of point dipoles is determined by the polarization autocorrelation function [3]. In both cases the response function is related to the static longitudinal dielectric function $e_L(k)$. Generalization of the theoretical development in [3] for systems of rigid ions to systems of both point charges and dipoles leads to a relation between $e_L(k)$ and a response function determined by charge-charge, dipole-dipole and charge-dipole correlations. To our knowledge, this is the first time that this relation has been deduced. We describe below the main steps of the generalized development.

We will focus on the linear response to a weak external electric potential $d F^{ext}(\mathbf{r})$ of the charge and dipole-moment densities,

$$r_q(\mathbf{r}) = \sum_{i=1}^{N} q_i d(\mathbf{r} - \mathbf{r}_i) \quad \text{and} \quad \mathbf{M}(\mathbf{r}) = \sum_{i=1}^{N} \mathbf{p}_i d(\mathbf{r} - \mathbf{r}_i), \tag{3}$$

respectively. In the absence of an external perturbation their averages vanish. The variation in the potential energy of equation (2) as a result of the external potential is

$$dU = \sum_{i=1}^{N} q_i dF^{ext}(\mathbf{r}_i) - \sum_{i=1}^{N} \mathbf{p}_i \cdot d\mathbf{E}^{ext}(\mathbf{r}_i), \tag{4}$$

where $d\mathbf{E}^{ext}(\mathbf{r}) = -\tilde{\mathbf{N}} dF^{ext}(\mathbf{r})$. The values of the dipole moments also change ($d\mathbf{p}_i \neq 0$) but it does not add any new term in equation (4) because $\tilde{\mathbf{N}}_{\mathbf{p}_i} U = \mathbf{0}$ [7]. Then, the variation in the grand partition function $\mathbf{X}$

$$d\mathbf{X} = \int \frac{d\mathbf{X}^q}{dF^{ext}(\mathbf{r}')} dF^{ext}(\mathbf{r}')d\mathbf{r}' + \int \frac{d\mathbf{X}^p}{d\mathbf{E}^{ext}(\mathbf{r}')} d\mathbf{E}^{ext}(\mathbf{r}')d\mathbf{r}' \tag{5}$$

with $$\frac{d\mathbf{X}^q}{dF^{ext}(\mathbf{r}')} = -b\mathbf{X}\langle r_q(\mathbf{r}')\rangle \quad \text{and} \quad \frac{d\mathbf{X}^p}{d\mathbf{E}^{ext}(\mathbf{r}')} = b\mathbf{X}\langle \mathbf{M}(\mathbf{r}')\rangle, \tag{6}$$

where $b = 1/k_B T$ with $k_B$ the Boltzmann constant and $T$ the temperature, and $\langle ... \rangle$ denotes the statistical average. From equations (6) and after long manipulation it is found that the Fourier components of the variation in $\langle r_q(\mathbf{r})\rangle$ and $\langle \mathbf{M}(\mathbf{r})\rangle$ are

$$d\langle r_q(\mathbf{k})\rangle = (\beta/V)[-\langle r_q(\mathbf{k})r_q(-\mathbf{k})\rangle d F^{\text{ext}}(\mathbf{k}) + \langle r_q(\mathbf{k})\mathbf{M}(-\mathbf{k})\rangle \cdot d\mathbf{E}^{\text{ext}}(\mathbf{k})] \qquad (7)$$

and
$$d\langle \mathbf{M}(\mathbf{k})\rangle = (\beta/V)[-\langle r_q(-\mathbf{k})\mathbf{M}(\mathbf{k})\rangle dF^{\text{ext}}(\mathbf{k}) + \langle \mathbf{M}(\mathbf{k})\mathbf{M}(-\mathbf{k})\rangle \cdot d\mathbf{E}^{\text{ext}}(\mathbf{k})], \qquad (8)$$

where
$$r_q(\mathbf{k}) = \sum_{i=1}^{N} q_i \exp(-i\mathbf{k}\cdot\mathbf{r}_i) \quad \text{and} \quad \mathbf{M}(\mathbf{k}) = \sum_{i=1}^{N} \mathbf{p}_i \exp(-i\mathbf{k}\cdot\mathbf{r}_i), \qquad (9)$$

and their correlations are computed in absence of the external field. Taking into account the results in [8], the static longitudinal dielectric function can be written as

$$\frac{1}{\varepsilon_L(k)} = 1 - c(k) = 1 + \frac{d\langle r_q(\mathbf{k})\rangle - i\mathbf{k}\cdot d\langle \mathbf{M}(\mathbf{k})\rangle}{dr^{\text{ext}}(\mathbf{k})}, \qquad (10)$$

where $c(k)$ is the dielectric response function and, from Poisson's equation, $dr^{\text{ext}}(k) = (k^2/4\pi)dF^{\text{ext}}$. If we put equations (7) and (8) in equation (10) it is found that $c(k)$ can be written as

$$c(k) = c_{qq}(k) + c_{pp}(k) + c_{qp}(k) \qquad (11)$$

where
$$c_{qq}(k) = 4\pi\beta\rho_N \frac{\langle r_q(\mathbf{k})r_q(-\mathbf{k})\rangle}{Nk^2}, \qquad (11a)$$

$$c_{pp}(k) = 4\pi\beta\rho_N \frac{\langle [\mathbf{k}\cdot\mathbf{M}(\mathbf{k})][\mathbf{k}\cdot\mathbf{M}(-\mathbf{k})]\rangle}{Nk^2}, \qquad (11b)$$

and
$$c_{qp}(k) = i4\pi\beta\rho_N \frac{\langle r_q(\mathbf{k})[\mathbf{k}\cdot\mathbf{M}(-\mathbf{k})]\rangle - \langle r_q(-\mathbf{k})[\mathbf{k}\cdot\mathbf{M}(\mathbf{k})]\rangle}{Nk^2}, \qquad (11c)$$

with $\rho_N = N/V$. Notice that $c_{qp}(k)$ includes the imaginary unity i. Nevertheless $c_{qp}(k)$, as well as $c_{qq}(k)$ and $c_{pp}(k)$, is a real number. The above equations yield the already known results for rigid ion systems (without dipole moments) or point dipole systems (without single charges) given in [3].

At this point it is worth noting that the assumption of perfect screening in conducting fluids, i.e., $\lim_{\mathbf{k}\to 0}[dr^{\text{ext}}(\mathbf{k}) + d\langle r_q(\mathbf{k})\rangle] = 0$, for rigid ion models yields the long-wavelength divergence $\varepsilon_L(k\to 0) = \infty$, from which it is derived that

$$\lim_{k\to 0} S_{ZZ}^{\text{RIM}}(k) = \frac{1}{k_D^2}k^2 \qquad (12)$$

where $S_{ZZ}(k) = (z^2e^2)^{-1}\langle r_q(\mathbf{k})r_q(-\mathbf{k})\rangle/N$ is the charge static structure factor and $k_D^2 = 4\pi z^2 e^2 \rho_N \beta$ is the square of the Debye wave number. The factor $1/z^2 = N/\Sigma_i z_i^2$ ensures that $S_{ZZ}(k)$ approaches unity as $k\to\infty$. However, for PIM it is derived

$$\lim_{k\to 0} S_{ZZ}^{\text{PIM}}(k) = \frac{k^2}{k_D^2}[1 - \tfrac{1}{2}\lim_{k\to 0} c_{qp}(k)]. \qquad (13)$$

## 4. Results

In the present work we carried out MD simulations over $3\times 10^5$ steps, with a time step of $5\times 10^{-15}$ s, using 1000 ions at $T = 923$ K and $\rho_N = 0.0281$ Å$^{-3}$. Computational details are described in [7]. The MD results of the static coherent structure factor for RIM and PIM are compared with experimental data [9] in figure 1. These results are discussed in [1]. We remark that the experimental prepeak at about $k_1 = 1$ Å$^{-1}$, which has been related to the inhomogeneous spatial distribution of cations [1,10], is reproduced only by the PIM. This prepeak becomes a shoulder for the PIM $S_{ZZ}(k)$ shown in figure 2. Both RIM and PIM $S_{ZZ}(k)$ exhibit a pronounced peak at $k_M = 1.7$ Å$^{-1}$, a clear sign of charge ordering, but present a different behavior at low $k$. The limits derived in equations (12) and (13) are in agreement with MD results (see the inset in figure 2) and account for this difference.

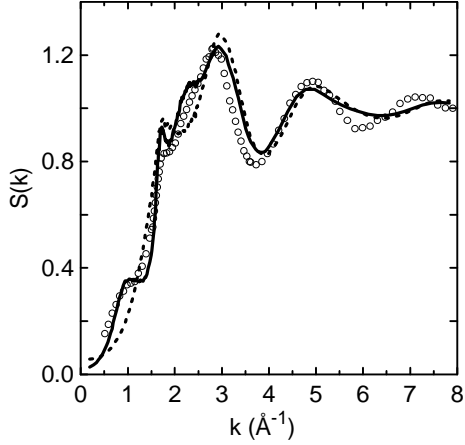 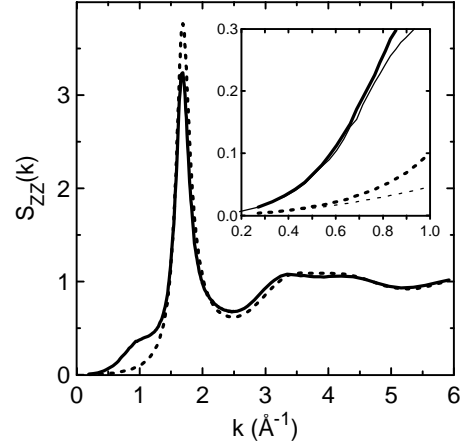

**Figure 1.** Static coherent structure factor from experimental data (open circles) [9] and MD results using the RIM (dotted line) and the PIM (solid line).

**Figure 2.** MD static charge structure factor $S_{ZZ}(k)$ for RIM (dotted line) and PIM (solid line). The inset shows $S_{ZZ}(k)$ at low $k$'s, as well as their limit equations (12) and (13) (thin lines).

The MD results of $e_L(k)$ are shown in figure 3. The $e_L(k)$ for RIM is similar to that found from neutron diffraction data of molten NaCl taking into account only equation (11a) [11] and from simulations of a rigid ion model of liquid alkali metals [12]. It diverges at $k = 0$ as it is expected from equation (12), has a maximum at $k_M$, and is negative in a wide $k$ region up to 4.6 Å$^{-1}$ where a second divergence is observed. At higher $k$'s it is positive and decreases monotonically to 1, the correct limit for rigid ion models [4]. On the other hand, the $e_L(k)$ for PIM starts from a positive value and increases to ∞ at about $k_1$. This result resembles those from simulations of dipolar liquids models [4,5]. As in the RIM case, the $e_L(k)$ for PIM has a relative maximum at about $k_M$, but remains negative as $k \to \infty$ as it has also been observed for a dipolar liquid in [5]. This $k \to \infty$ limit is due to the dipole-dipole correlation (see below). If only the $c_{qq}(k)$ term was considered to calculate the $e_L(k)$ for PIM, there would be a divergence at 4.6 Å$^{-1}$, but not at $k_1$.

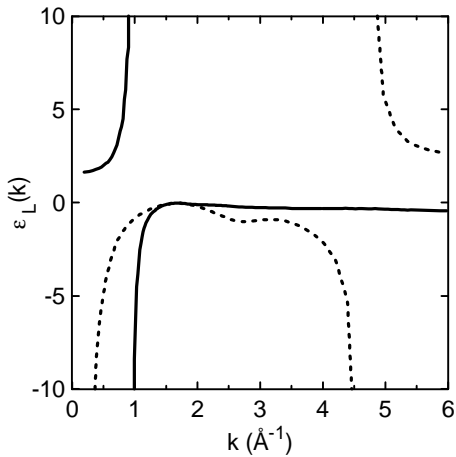

**Figure 3.** Static dielectric longitudinal function $e_L(k)$ for RIM (dotted line) and PIM (solid line).

More insight is gained by looking at the response function $c(k)$ and their contributions plotted in figures 4 and 5. $c(k)$ relates the screened total potential $F(k)$ to the external potential $dF^{ext}(k)$ by means of $c(k) = 1 - F(k)/dF^{ext}(k)$ [4]. Then, $c(k) > 1$ implies that $F(k)$ and $dF^{ext}(k)$ have an opposite sign (*overscreening effect*). The *overresponse* is maximum at $k_M$, where $c(k) \gg 1$ due to the $c_{qq}(k)$ contribution. This means that the bound charge in the molten salt, whose short-range periodicity is $\sim 2\pi/(k_M)$, screens the $k_M$ component of the external potential. Figure 5 shows that the different behavior at larger $k$'s of the $c(k)$ for RIM and PIM, and thus of their $e_L(k)$, is due to $c_{pp}(k)$, which does not tend to zero as $k\rightarrow\infty$. It is worth noting that the PIM $c(k)$ does not present any feature at $k_1$ in spite of the relative maxima of $c_{qq}(k)$ and $c_{pp}(k)$, and the minimum of $c_{qp}(k)$, at this wave number.

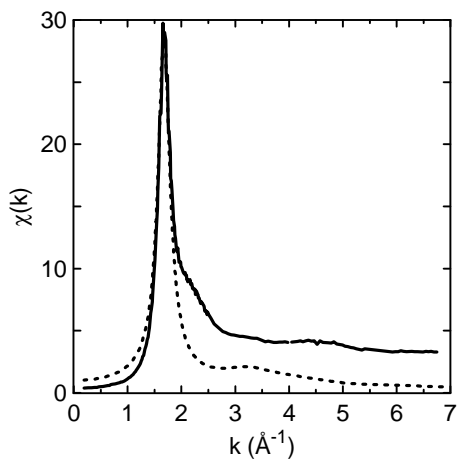 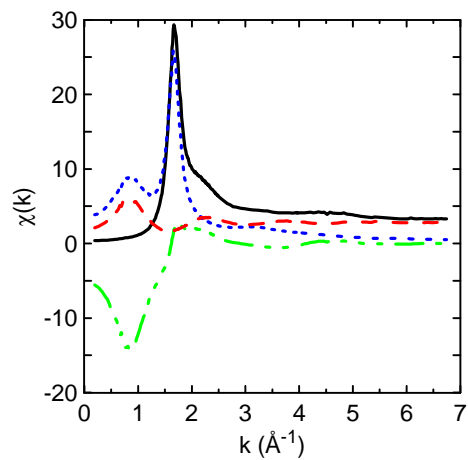

**Figure 4.** Response function $c(k)$ for RIM (dotted line) and PIM (solid line).

**Figure 5.** Contributions to PIM $c(k)$ (solid line): $c_{qq}(k)$ (dotted line), $c_{pp}(k)$ (dashed line), and $c_{qp}(k)$ (dash-3-dots line).

**Acknowledgments**
This work was supported by DGICYT of Spain (FIS2006-12436-C02-01), the DURSI of the Generalitat of Catalonia (2005SGR-00779) and European Union FEDER funds (UNPC-E015). One of us (V.B.) thanks the Ministry of Education of Spain for the FPU Grant No. AP2003-3408.